\documentclass[12pt]{article}

\usepackage{times,color}
\usepackage{amsmath}
\usepackage{subfigure,epsfig,graphicx,epstopdf}
\usepackage[colorlinks]{hyperref}
\hypersetup{linkcolor = {blue},	citecolor = {blue},	urlcolor = {blue}}
\usepackage{cite}

\newenvironment{sciabstract}{%
\begin{quote} \bf}
{\end{quote}}

\newcounter{lastnote}

\begin{document}

\title{Error-Corrected Eternal Lifetime Storage} 

\author{Jie Ma$^{1,2}$,  Chu-Han Wang$^{1,2}$, Xiao-Yun Xu$^{1,2}$,  Chang-Kun Shi$^{5}$, Tian-Yu Zhang$^{1,2}$,\\ Ke Cheng$^{1,2 \ast}$, Li Zhan$^{3 \ast}$, Xian-Min Jin$^{1,2,4,5\ast}$ \\
\normalsize{$^1$Center for Integrated Quantum Information Technologies (IQIT), School of Physics and Astronomy}\\
\normalsize{and State Key Laboratory of Advanced Optical Communication Systems and Networks,} \\
\normalsize{Shanghai Jiao Tong University, Shanghai 200240, China}\\
\normalsize{$^2$Hefei National Laboratory, Hefei 230088, China}\\ 
\normalsize{$^3$State Key Laboratory of Advanced Optical Communication Systems and Networks,}\\
\normalsize{School of Physics and Astronomy, Shanghai Jiao Tong
University, Shanghai, 200240, China}\\
\normalsize{$^4$TuringQ Co., Ltd., Shanghai 200240, China}\\
\normalsize{$^5$Chip Hub for Integrated Photonics Xplore (CHIPX),}\\
\normalsize{Shanghai Jiao Tong University, Wuxi 214000, China}\\
\normalsize{$^\ast$E-mail: chengkek@sjtu.edu.cn}\\
\normalsize{$^\ast$E-mail: lizhan@sjtu.edu.cn}\\
\normalsize{$^\ast$E-mail: xianmin.jin@sjtu.edu.cn}
}

\date{}

\baselineskip24pt

\maketitle

\begin{sciabstract}
In the information explosion era, the demand for high-density stable storage technologies is soaring. Multi-dimensional optical storage with femtosecond laser writing offers a potential solution for massive data storage. However, equipment instability and reduced voxel resolution inevitably lead to data errors. Here, we propose and demonstrate a paradigm exemplifying high-fidelity eternal lifetime optical storage enabled by error correction mechanism. We increase information density by reducing voxel size and spacing. Leveraging deep learning methods, we achieve 8-bit voxel encoding and a storage capacity of 2.15 $Tb/disc$. We implement the Reed-Solomon(RS) algorithm for error-free data recovery and get the trade-off between the storage capacity and the redundancy length. Our storage paradigm takes advantage of error-correcting codes, together with permanent information storage capabilities of extremely stable fused silica, marking a significant advancement for recording massive data to the application level and making it possible to faithfully record the information generated in human civilization.
\\
\end{sciabstract}

\section*{Introduction}
\noindent With the rapid development of Internet technology, the total amount of global data is projected to reach 175 ZB by 2025, according to a report released by the International Data Corporation\cite{rydning2018digitization,gu2014optical}. Most of them are cold data that  necessitate prolonged and stable archival storage solutions\cite{levandoski2013identifying}. Nowadays, data centers based on hard disk drive arrays or solid-state drive are the predominant methods used for storing massive data\cite{gu2014optical}. However, these methods have limitations, including a limited service life, sensitivity to environmental factors, and high energy consumption during data migration\cite{fleischer2020cooling}. The need for high data capacity, enhanced security, and long-lifetime storage has become an urgent issue to be addressed. Optical storage technology is a novel selection for information storage due to its unique advantages including long lifespan, low energy consumption and large capacity\cite{gu2014optical,zhang2014seemingly,gu2010road}. Various storage methods have been proposed like multi-wavelength multi-stage optical storage\cite{stiller2017cascaded,seddighian2008optical}, beam super-resolution optical storage\cite{meiling2019research,wei2013dynamic}, and holographic optical storage\cite{mansuripur1997principles,okamoto2011holographic,hvilsted2009volume}. Compared to these methods, optical storage based on fused silica offers a distinct advantage by confining the interaction of light and matter to nanoscale and enabling stable mass storage\cite{lei2023ultrafast,glezer1996three,qiu1998three,zhang2016eternal,bhardwaj2006optically}.

The femtosecond direct writing process in fused silica induces modifications in the focal area, resulting in the formation of a stable anisotropic nanograting structure\cite{shimotsuma2003self,shimotsuma2010ultrafast}. This structure exhibits birefringent properties and allows for permanent information storage at room temperature\cite{zhang2014seemingly}. Recent advancements have concentrated on improving storage modes and capacities, such as the utilization of novel structures to enable multi-layer optical storage\cite{sakakura2020ultralow,wang2022100} and the attainment of storage densities reaching 1.64 $Tbit/disc$\cite{lei2021high}. Nevertheless, the adhesion between adjacent voxels will pose challenges in extracting data, and higher coding bits may lead to indistinguishable colors to image recognition errors\cite{wang2022100,yang1999interlayer,yi1995statistical,gao20214d}. Machine learning helps achieve high-resolution and high-accuracy data readout and can be applied to optical data storage in nanophotonics\cite{lamon2021nanophotonics}. However, as the encoding information capacity increases and the voxel spacing decreases, deep learning methods face application-level bottlenecks that prevent them from achieving 100\%  accuracy. Furthermore, the occasional jitter of femtosecond laser and the inhomogeneity of the storage media may result in missing or incorrect written data, thereby impacting the accuracy of data storage\cite{lei2023ultrafast}.  In general, the storage process is affected by various factors such as noise, distortion, interference and even extreme cases such as disc damage, which may lead to data loss\cite{van2006cancellation,dong2011estimating,wang2011error}. These information losses and errors caused by inherent system errors or accidental factors cannot be achieved through deep learning to improve recognition accuracy. Consequently, to ensure error-free access to data, it becomes imperative to introduce an error correction mechanism that can effectively address and rectify these data imperfections.

Here, we report an experimental optical storage paradigm with high-fidelity and large capacity in silica using error correction mechanism. We enhance information density by using a high numerical aperture object lens to form smaller voxels and utilize TPH-YOLOv5 and VGG16 algorithms to assist in higher-bit encoding\cite{zhang2022low}. Ultimately, we successfully achieve encoding and decoding in 64 slow axis azimuth angles and 4 retardance amounts, resulting in an accuracy of 92.49$\%$ with the presence of accidental errors. To ensure that the data can be retrieved accurately after reading, we employ RS (255, 218) which indicates that the encoder takes 218 (8-bit, 1 byte = 8 bits) information symbols and 35 parity symbols to construct a 255 byte information sequence and verify the reliability for various error types. Each benchmark information sequence can be utilized as a reliable data sector architecture unit to achieve data storage with high fault tolerance coding. Our approach achieves high-fidelity data storage with a capacity of up to 2.15 $Tb/disc$, making a significant contribution to large-scale data recovery and future storage technology.

\section*{Results}

\begin{figure*}
\centering
\includegraphics[width=1 \columnwidth]{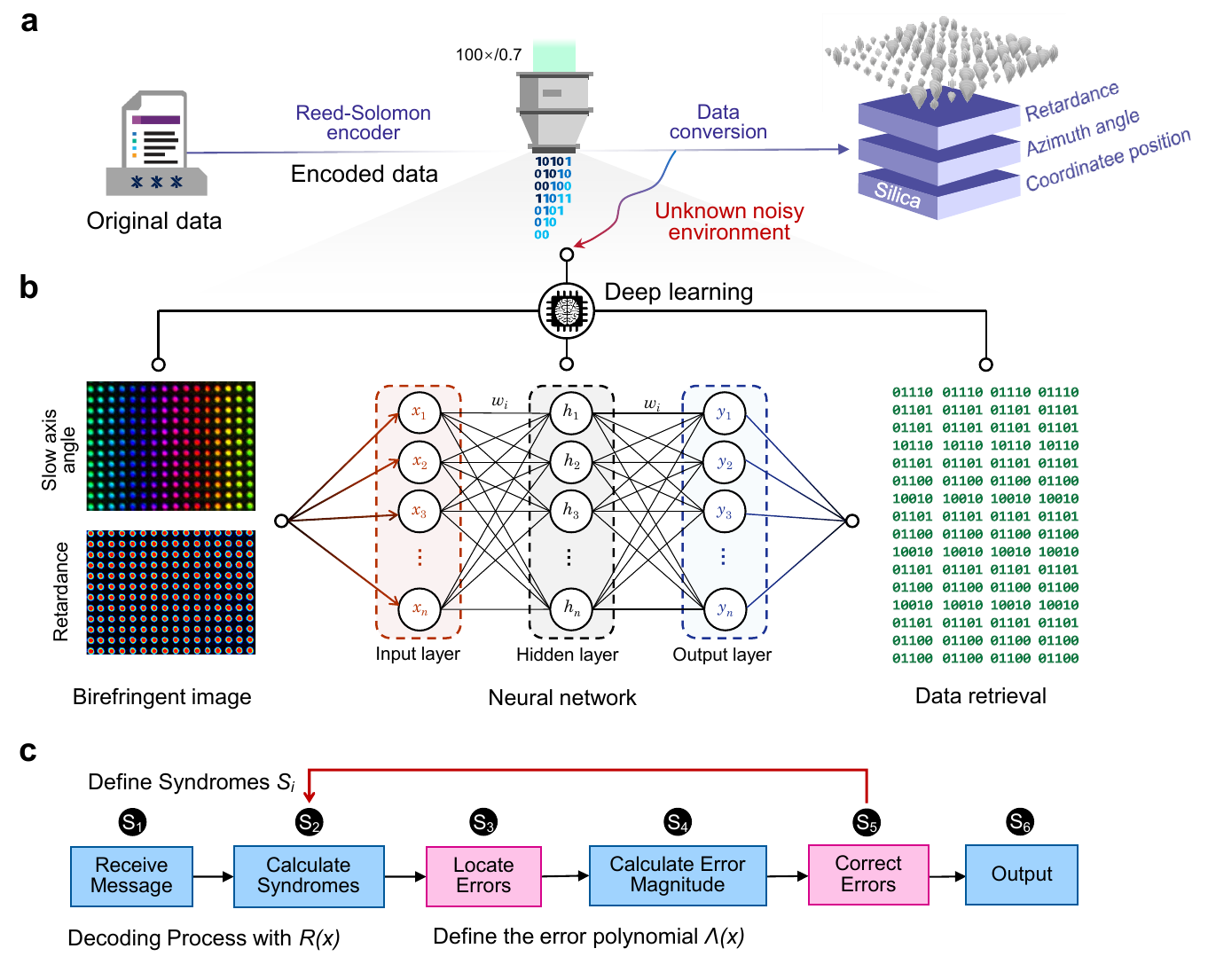}
\caption{\textbf{Schematic illustration of a data storage system with error correction based on Reed-Solomon codes. } \textbf{(a)} The original data is based on RS codes redundancy code addition and data conversion. \textbf{(b)} Use the TPH-YOLOv5 algorithm to locate the data, and use the VGG16 algorithm to classify and label the collected birefringence information, achieving data recovery. \textbf{(c)} The process of decoding using Reed-Solomon error correction algorithm. }
\label{fig1}
\end{figure*}

The overall process of implementing five-dimensional optical data storage is illustrated in Fig.\ref{fig1}(a). We first transform the information into binary data and then integrate redundant data by utilizing RS codes\cite{wicker1999reed}. We further transform the data into five-dimensional information consisting of three-dimensional coordinate information, femtosecond laser polarization and power information (Section 1, Supporting information). We choose fused silica as the storage medium because it has a high Young's modulus, wear resistance, and thermal stability, which allows for permanent data storage\cite{zheng2007surface,penn2006frequency,xiao2021models}. When the femtosecond laser directly writes inside fused silica, material denaturation occurs due to the interaction between laser and material, leading to formation of nanostructures under specific parameters\cite{sudrie1999writing,mills2002embedded,shimotsuma2003self,glezer1996three}. Through the high-precision birefringence microscopy system, the stored birefringence information of anisotropic nanostructure arrays can be acquired. Based on the retardance and slow axis azimuth information reconstruted by specific algorithm, the data can be obtained using deep learning method as shown in Fig.\ref{fig1}(b). The retrieved data encompasses both the original dataset and redundant information introduced through RS codes rules. We subsequently employ the decoding algorithm displayed in Fig.\ref{fig1}(c) to rectify potential data errors and yield the original data as the final output.

Increasing storage capacity can improve storage efficiency, which determines the feasibility of data storage and archiving and is one of the important considerations for optical storage. Smaller nanograting structures can be formed to enable more voxels to be written in a unit volume so as to increase storage capacity. However, the laser is limited by the optical diffraction limit, and the minimum focus point size is given by $ d = 0.61 \lambda /NA $, where $d$ represents the resolution distance of the objective lens, $ \lambda $ is the laser wavelength, and $ NA $ is the numerical aperture of the objective lens\cite{fischer2013three,zheludev2008diffraction,latychevskaia2019lateral}. We employ a 100X, 0.7NA high numerical aperture objective to reduce the laser focus and the theoretical size of the voxel formed is 0.45 microns. 
Reducing the voxel spacing can enhance storage capacity, as more storage units can be formed in a unit space. However, the nanograting’s longitudinal and lateral dimensions pose a challenge: small spacing causes crosstalk between different nanogratings, leading to recognition errors and affecting the accuracy of readout.\cite{taylor2008applications}. Balancing appropriate voxel spacing and ensuring accurate data readout can become an applicable storage model. Later, we discuss the incorporation of error correction mechanisms as a means to achieve the dual objectives of capacity and data fidelity. Higher-bits encoding in retardance and slow axis azimuth of individual voxel, which are determined by the intensity and polarizaiton angle of the femtosecond laser, can effectively increase the capacity of recorded information. We use the retardance and the slow axis azimuth information to encode a single voxel with 5 bits ($2^{5}=32$ types, a total of 32 states corresponding to the retardance amount or slow azimuth angle), 6 bits, 7 bits, and 8 bits respectively. For data classification and training, we adopt the VGG16 algorithm for high-precision analysis, and a self-trained classifier is used to improve the classification ability of confusing categories. Finally, we test on random data to evaluate its generalization performance and accuracy. Additional details about recognition and classification using TPH-YOLOv5 and VGG16 algorithm are elucidated in Methods.

\begin{figure*}
\centering
\includegraphics[width=1 \columnwidth]{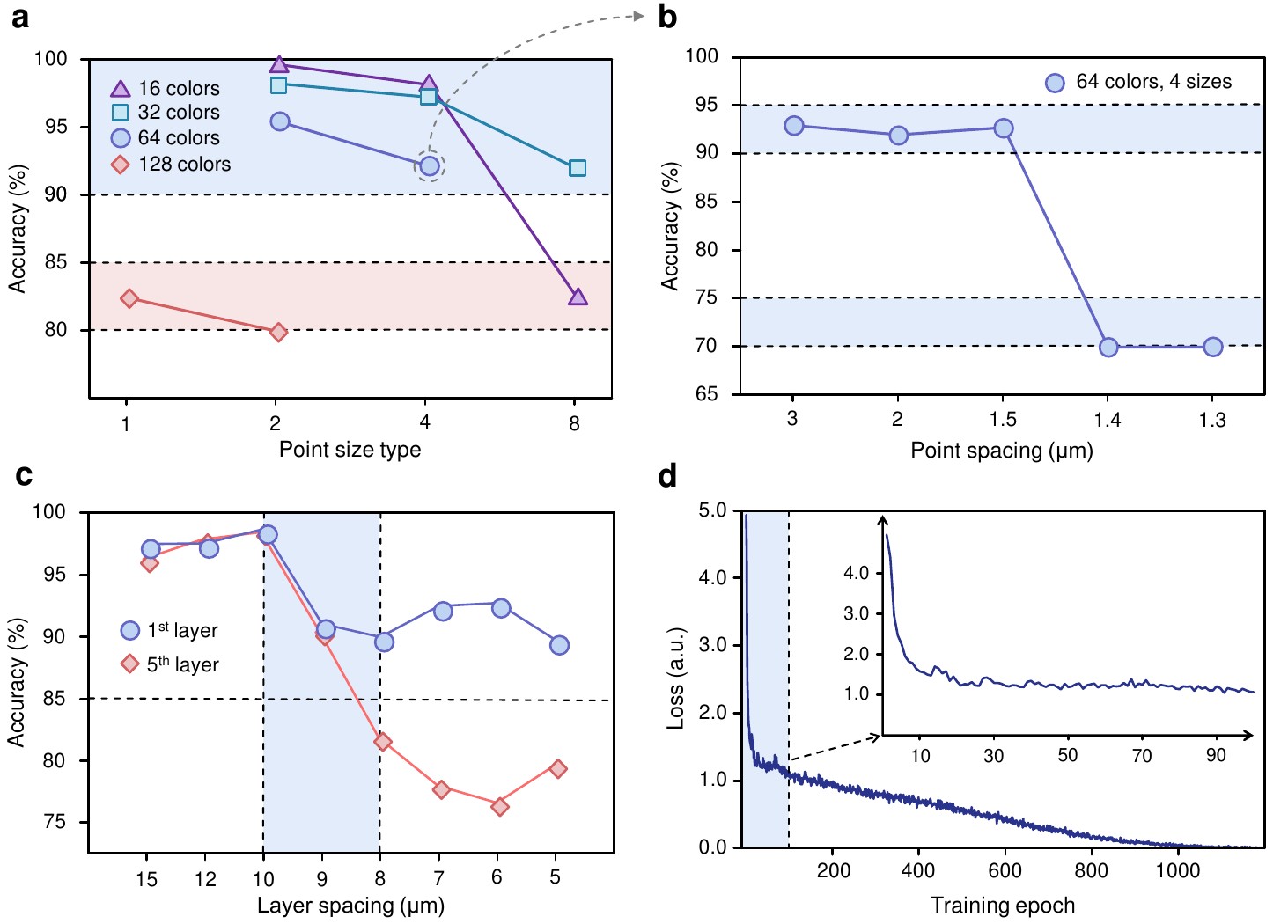}
\caption{\textbf{The modification of various parameters is for the accuracy of data recognition. } \textbf{(a)} The recognition accuracy of different point sizes under 16, 32, 64, 128 colors. \textbf{(b)} The recognition accuracy of points of 64 colors and 4 sizes as the point spacing changes. \textbf{(c)} The recognition accuracy of the middle point of the multi-layer data changes with the layer spacing (Taking the first layer and the fifth layer data as an example).
\textbf{(d)} Loss changes with Epoch during VGG16 training.
} 
\label{fig2}
\end{figure*}

The accuracy of points with different encoding bits is displayed in Fig.\ref{fig2}(a). The accuracy rate can reach as high as 95.85$\%$ and 92.50$\%$ at 7 bits (2 sizes and 64 color classifications) and 8 bits (4 sizes and 64 color classifications) respectively. Nevertheless, it is observed that the accuracy rates for 128 colors are relatively low, with only 82.65$\%$ and 80.17$\%$ accuracy for one and two size classifications respectively. Increasing the retardance classification leads to blurred point sizes and a sharp decline in accuracy under 8 sizes. Although a higher retardance threshold can be chosen to increase difference resolution, single voxel structure will be enlarged, leading to crosstalk between layers and impacting storage density. Therefore, we have selected 4 types of sizes as coding standard for the retardance dimension, i.e., 64 types of slow axis azimuth classification, 4 types of retardance and a total of 8 bits. We verify the accuracy of the device under real data in consideration of the large points that have a greater impact on the spacing at the minimum point spacing attempt. We evaluate the classification accuracy when the voxel spacing gradually decreased from 3 microns for 64 slow azimuth and 4 retardance classifications information encoding as displayed in Fig.\ref{fig2}(b). A high accuracy rate of 92.92$\%$ is still achieved when the point spacing is 1.5 microns. However, the accuracy rate decreased sharply to 70.19$\%$ when the spacing decreased to 1.4 microns, indicating that further reducing the point spacing under the current experimental parameters would significantly increase the mutual interference between points in reasonable range.

For multilayer spacing, we write ten layers of data under 64 azimuth classifications and the largest retardance, and reduce the layer spacing to verify the accuracy of recognition after training. Considering the loss of laser energy after passing through the medium (The absorption and scattering loss of the laser in the medium leads to different modification degrees of materials at different working distances.), we modify the laser power to compensate for the loss of laser light after passing through the medium\cite{kogelnik1965propagation,mccall1969self}. As shown in Fig.\ref{fig2}(c), the accuracy rate was higher than 90$\%$ for both the first layer and the fifth layer data when the layer spacing was 9 microns, but there was a significant decline. Further reduction of layer spacing resulted in an instantaneous decrease in the accuracy rate, indicating that the longitudinal size of each voxel results in strong layer-to-layer crosstalk. When the point spacing is 1.5 microns and the layer spacing is 10 microns, the loss curve during training for 8-bit classification is shown in Fig.\ref{fig2}(d).

\begin{figure*}
\centering
\includegraphics[width=1 \columnwidth]{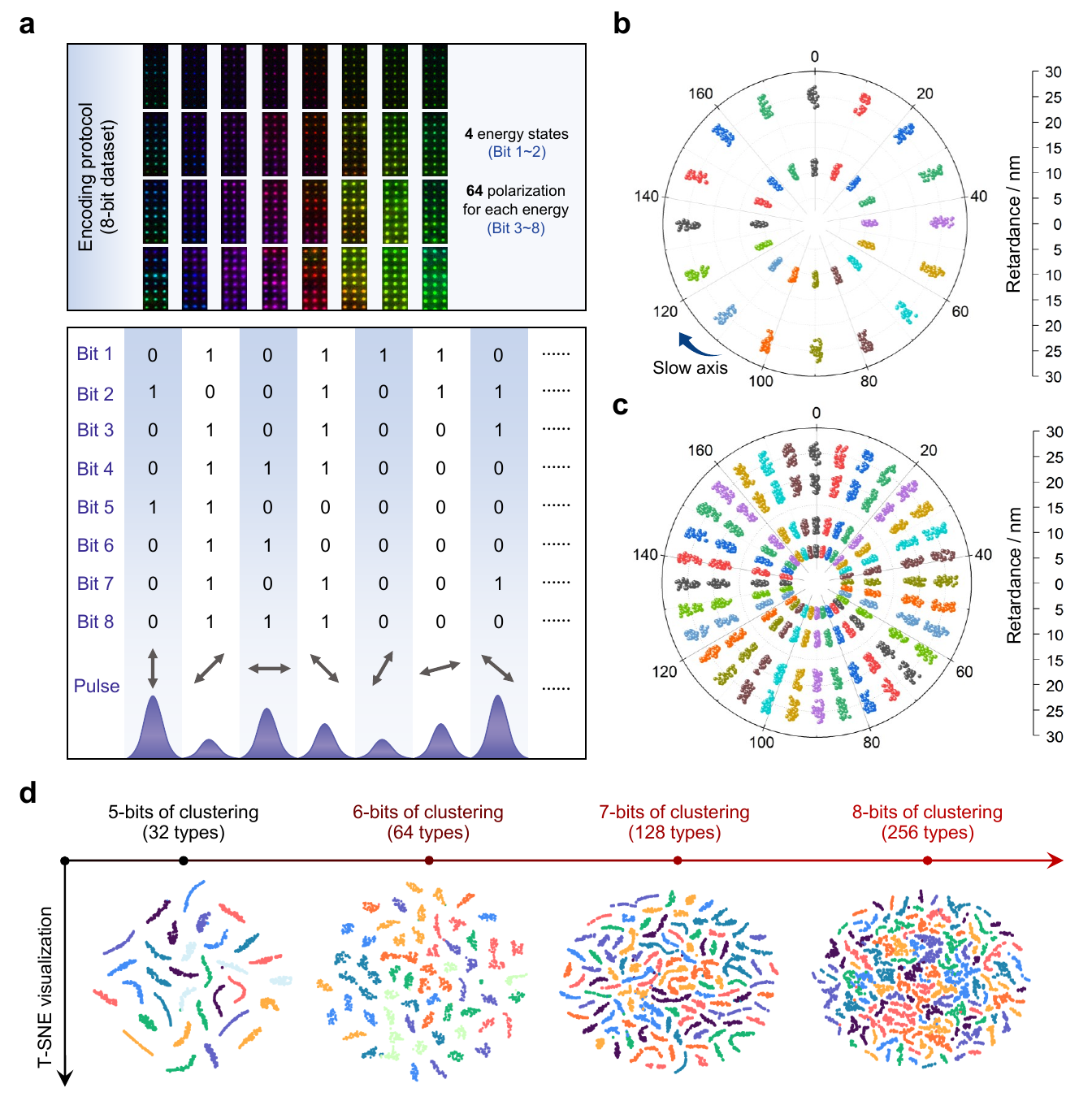}
\caption{\textbf{Data coding and cross-validation based data classification validity analysis.} \textbf{(a)} Illustration of the encoding protocol. The bit 1 and bit 2 represent 4 different laser energy states. At the same time, the bit 3-bit 8 represent 64 different laser polarization for each energy. The bottom half of figure 3(a) shows the laser pulse energy and polarization under some coding combinations.
\textbf{(b)} and \textbf{(c)} use clustering methods to obtain polar maps of retardance and slow axis azimuth distributions for 32 and 128 different voxels, respectively. Among them, there are 2 types of retardance and 16 types of azimuth angle in \textbf{(b)}; there are 4 types of retardance and 32 types of azimuth angle in \textbf{(c)}. \textbf{(d)} t-SNE clustering diagram based on 32, 64, 128 and 256 types of data.}
\label{fig3}
\end{figure*}

In subsequent experiments, we utilize an 8-bit encoding scheme to represent a voxel, corresponding to a total of 256 states. The rules between label samples and encoding are shown in Fig.\ref{fig3}(a). The polar coordinate plots representing the retardance and slow axis azimuth distribution of different voxels, categorized under 32 and 128 classes using cluster analysis, are depicted in Fig.\ref{fig3}(b) and Fig.\ref{fig3}(c) respectively. Employing t-distributed stochastic neighbor embedding (t-SNE) dimensionality reduction technique\cite{belkina2019automated}, the visual classification is performed on the distribution of different numbers of species, different colors represent different classification types, as illustrated in Fig.\ref{fig3}(d). The cluster analyses of 32, 64 and 128 categories have obvious classification, and the misjudgment between categories is relatively small. For the cluster analysis under 256 categories, the data can still maintain a high classification accuracy, except that individual points overlap each other and cause data misidentification. This shows that the encoding of 64 kinds of slow azimuth angle and 4 kinds of retardance categories is feasible.

In addition to the data reading errors mentioned earlier resulting from increased density, data errors during laser direct writing can also be caused by equipment and environmental factors. For instance, deviations during data writing due to inhomogeneities in the modified substrate can significantly undermine data accuracy. Depending on the potential scenarios, error types can be divided into scatter errors, chain errors and block errors\cite{kanal1978models}. Mechanical fatigue-induced polarizer rotation errors exert a profound impact on system stability, precipitating the emergence of scatter errors. Chain errors predominantly originate from irregularities in the internal materials during the forging process of the storage medium. To achieve error-free information recovery, we rely on the implementation of error-correcting codes to fulfill this objective.

The RS codes are linear error correction codes that are widely used in data storage (such as CD, DVD and Blu-ray disc) and data information transmission\cite{shrivastava2013error}. Its encoding process involves linearly operating the binary data sequence to be stored with the user-defined generating polynomial in the Galois Field $GF (2 ^ {m})$ to produce a binary sequence containing both real data and redundant data. The decoding process entails restoring the read information (the collected data) to binary information, computing the adjoint polynomial $ S_j $ from the acceptance polynomial $R(x)$, locating the error location and information. Then, we can correct the error data, and finally output the data\cite{wicker1999reed}.

\begin{figure*}
\centering
\includegraphics[width=1 \columnwidth]{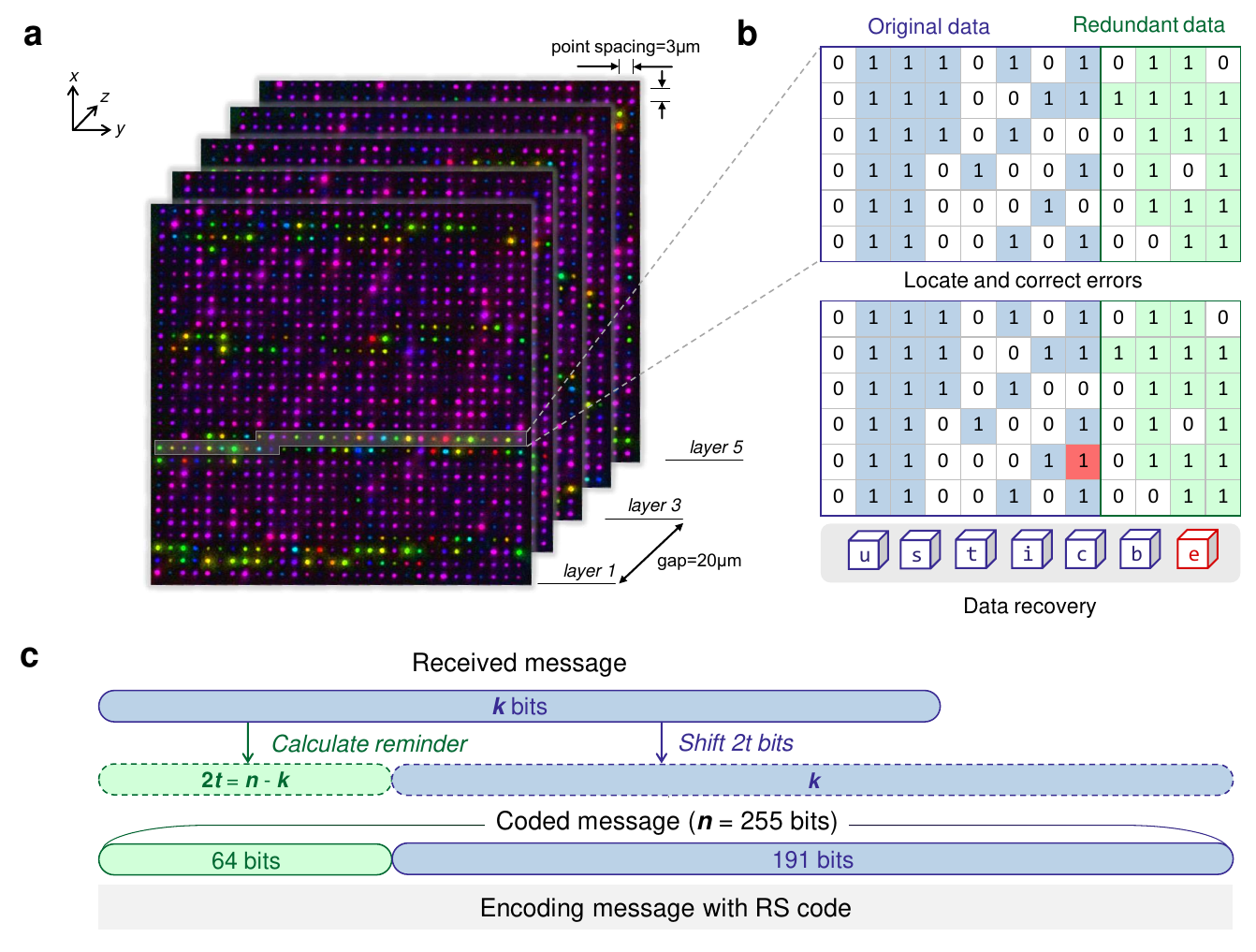}
\caption{\textbf{Utilizing error correction codes to achieve high-fidelity recovery of data. } \textbf{(a)} A multi-layer representation of Martin Luther King's article “I have a dream” measured using a birefringence microscopy system with a layer spacing of 20 $ \mu m $ and a point spacing of 3 $ \mu m $.
\textbf{(b)}
Restore some of the data selected in Fig.4(a) according to the corresponding rules, including original data and redundant data, and use error correction algorithms for error recovery.
\textbf{(c)}
Structure of RS codes under 255 bits. The k bits part is the original data part (blue) and the 2t bits part is the redundant data part (green), which has t bits of error correction capability. The figure shows the structure of the RS (255,191) error correction codes, in which the total length of the data string is 256 bits, and a maximum of 32 bits of error information can be recovered.
}
\label{fig4}
\end{figure*}

The multilayer real data of the first five layers as an example are shown in Fig.\ref{fig4}(a). We selected 3 microns point spacing and 10 microns layer spacing for display so as to show the data storage results more clearly. Error correction for the existing error data is shown in Fig.\ref{fig4}(b), where the blue box contains raw data, and the red box contains redundant data. RS codes are used to achieve the positioning and error correction of data. The structure of RS codes is shown in Fig.\ref{fig4}(c). Using RS codes with varying redundancy length can realize good recovery for all three error types mentioned above. For general data, after using TPH-YOLOv5 for precise positioning and VGG16 algorithm for efficient data classification, error rate can be kept low\cite{yang2023deep,lecun2015deep}. For the analysis of the results obtained from a large number of data tests, the error rate is about 8$\%$. We chose the RS (255, 218) codes for error correction based on the error-correcting principle of the RS codes. Its error correction ability for data can just achieve perfect data recovery. During the storage process, we achieved a trade-off of a storage capacity of 2.15 $Tb$ and a redundancy length of 38 bits (Section 4, Supporting information). In this case, we have achieved a high-fidelity optical data storage with a storage density of 2.15 $Tb/disc$ in glass.

\section*{Conclusion}
In conclusion, we achieve high-fidelity optical storage with large capacity in silica. We use a high numerical aperture objective lens to reduce the size of the voxel and thereby further narrow the voxel interval. With the help of the TPH-YOLOv5 algorithm and the VGG16 algorithm, the single voxel information encoding can reach 8 bits, thus realizing a higher data storage density. With the help of RS codes, we achieve error-free recovery of data after processing by error correction algorithm, and then test the effectiveness of its error correction under different data error types. Finally, we get the selection of RS codes based on the existing conditions, and the trade-off between the storage capacity and the redundancy length is given according to the identification situation.

Near-field optical technology has been greatly developed, which is of great significance for avoiding the diffraction limit\cite{fischer2013three}. In the future, based on the existing research foundation, combined with near-field optical technology, the multi-dimensional optical storage of femtosecond lasers will be realized by using solid immersion high numerical aperture objective lenses such as oil immersion lenses. This will enable further breakthroughs in storage capacity \cite{lei2021high,tan2021photonic}. In addition, the development of nanomaterials technology has provided a new direction for the next generation of optical data storage systems. Optical storage technology based on nanomaterials may be able to achieve low-power and high-accuracy data reading\cite{gu2016nanomaterials,gao20214d}.In terms of point recognition accuracy, using advanced image recognition algorithms to obtain a more general model may open new prospects for the improvement of recognition accuracy. The error-correcting codes based data storage model proposed in this article provides a feasible solution for building large-scale storage systems. It can realize a sustainable cloud archive storage mechanism based on the properties of fused silica and make it posssible to faithfully and externally record all the information generated in mankind development.

\section*{Methods}
\noindent \textbf{Data arrays preparation and readout measurement} \\
The device model diagram for data writing and reading is shown in Supplementary materials (Section 2, Supporting information). We use femtosecond laser direct writing in fused silica to form anisotropic nanostructures with birefringence properties to store information.
Femtosecond laser (Light Conversion Ltd.) generated by a regenerative amplifier based on Yb:KGW laser medium is focused by an objective lens (100×, 0.7 NA) into fused silica. Notably, the power of the laser is locked at 0.5\% rms throughout the duration of the experiment, while the polarizer is maintained within a deviation range of ±0.01 degrees. The wavelength of the laser is 515 nm and the repetition rate is 1 MHz. We use a high-precision three-dimensional translation stage that ensures a fixed movement rate. The jitter of inclination angle and three-dimensional coordinates are 0.5 $\mu rad$ and ±50 nm respectively.

The image acquisition and program analysis readout system based on the electronically controlled liquid crystal polarization compensator can extract the data inside the fused silica with high precision. The system modulates the different orientation angles of elliptically polarized light so that it passes through the sample and the polarized light analyzer composed of a 1/4 wave plate and a polarizer to obtain linearly polarized light with different light intensities. Finally, the CCD is used to capture the optical information under different orientation angles, and the image restoration algorithm is used to calculate the corresponding retardance and slow azimuth angle information.
\\

\noindent \textbf{Recognition and classification using neural networks} \\
TPH-YOLOv5 is a one-stage object detection algorithm that provides a conspicuous advantage in terms of object recognition speed. Given the multi-dimensional encoding of retardance of data, there are multiple data points that require optical frame selection to achieve accurate detection of objects of varying sizes.

Firstly, we use a birefringent microscope to take a batch of data images with a resolution of 2048 $ \times $ 2048 pixels. In order to enhance the model's generalization ability and prevent potential overfitting issues during subsequent training, we employ data augmentation techniques such as random rotation, scale transformation, and HSV enhancement, based on the shape and color features of the data points. The optimizer used for training is AdamW, which can automatically adjust the learning rate and greatly improve the training speed. After getting the initial model, we use TPH-YOLOv5 to crop and extract small images of 32 $ \times $ 32 pixels with the same label from each large image as the training sets.

On account of the low accuracy and efficiency of TPH-YOLOv5 classifier, we pack small images by category and feed them into VGG16 for training. VGG16 performs well in neural network classification algorithm due to its three-layer fully connected architecture. In particular, it utilizes multiple consecutive 3 $ \times $ 3 convolution kernels to improve classification results over larger convolution kernels. By leveraging multilayer nonlinear layers, it can achieve a greater depth of network to learn more complex patterns and distinguish between different types of points accurately.

We use the Focal loss function to ensure the deep recognition of samples that are difficult to distinguish and continue to use the AdamW optimizer for training. For each type of sample, we have 1600 small images for training, while 1280 of them are used for training, and the remaining 320 points are used for result inspection to verify the effectiveness of VGG16 network training. Finally, we test a substantial amount of real data under the trained VGG16 neural network and compare it with the original data to obtain its recognition accuracy data. (The flow chart of the data recovery process is shown in Section 1, Supporting information)
\\

\noindent \textbf{High-fidelity data recovery based on Reed-Solomon algorithm} \\
In order to ensure the integrity of stored data and prevent errors, a preprocessing step is performed prior to the conversion of binary data into multi-dimensional data. Specifically, the RS error correction algorithm is used to add redundant codes. RS codes are form of low-rate channel coding that employ forward error correction to prevent data packet loss during network transmission. In the linear block error correction code, it has the best error correction ability and coding efficiency\cite{wicker1999reed}. Researchers have found wide application in current storage methods to correct random errors and burst errors and improve storage stability.

In RS codes, the maximum error information bits that can be corrected are determined by $ t = (n - k)/2 $. However, the introduction of an erasure correction algorithm may lead to an increase in the data redundancy information. To address this issue, when missing information is encountered during data reading, the missing bits are randomly filled with 0 or 1, converting them from missing information to error information. The RS error correction rules are then used to correct these errors. During the decoding process, the position and information of errors are located and corrected through accepting the adjoint polynomials calculated by the polynomials to obtain output data.
\\

\noindent \textbf{Noise processing for high fidelity data restoration} \\
In the realm of data preservation and recovery, the presence of noise impedes the accurate retrieval of valuable information. In the process of data acquisition and recovery, the presence of impurities on the sample surface induces light scattering, thereby introducing scattering noise into the captured images. Additionally, the electronic and camera acquisition processes themselves give rise to Gaussian noise. 

In our work, we set a reasonable threshold for the background noise signal and employing a median filter to replace the intensity value of a specific pixel with the median intensity value of its neighboring pixels. This strategy serves to enhance the contrast and fidelity of the image data, enabling more accurate recovery of the desired information. Furthermore, The edge detection algorithm is used to isolate the pixels of the data points in the image, and the statistical information of adjacent pixels is used to optimize the filtering values of the relevant noise points to protect the edge details of the image.  Ultimately, through meticulous image cropping, the final information points can be obtained. To assess the accuracy of information point identification, appropriate parameters can be selected for binarization on the retardance information map, facilitating a comparison with the extracted information points.

\subsection*{Author contributions}
X.-M.J. conceived and supervised the project. J.M. performed the simulations, fabricated the storage chips, performed the experiment and analyzed the data. J.M. interpreted the data and wrote the paper, C.-H.W. optimized the figures and modified the text. All authors provided comments on this article. X.-Y.X. and C.-K. S. revised the manuscript and provided valuable comments. X.-M.J., L.Z. and K.C. supervised the entire project.

\subsection*{Acknowledgments}
This research is supported by the National Key R\&D Program of China (Grants No. 2019YFA03
08703, No. 2019YFA0706302, and No. 2017YFA0303700); National Natural Science Foundation of China (NSFC) (Grants No. 62235012, No. 11904299, No. 61734005, No. 11761141014, and No. 11690033, No. 12104299, and No. 12304342);  Innovation Program for Quantum Science and Technology (Grants No. 2021ZD0301500, and No. 2021ZD0300700); Science and Technology Commission of Shanghai Municipality (STCSM) (Grants No. 20JC1416300, No. 2019SHZDZX01, No.21ZR1432800, and No. 22QA1404600); Shanghai Municipal Education Commission (SMEC) (Grants No. 2017-01-07-00-02-E00049); China Postdoctoral Science Foundation (Grants No. 2020M671091, No. 2021M692094, No. 2022T150415). X.-M.J. acknowledges additional support from a Shanghai talent program and support from Zhiyuan Innovative Research Center of Shanghai Jiao Tong University.

\subsection*{Conflict of interests}
The authors declare no conflict of interests.

\subsection*{Data availability}
The data that support the findings of this study are available from the corresponding authors on reasonable request.

\bibliographystyle{unsrt}

\clearpage

\begin{thebibliography}{99}

\bibitem{rydning2018digitization} Reinsel, D., Gantz, J., Rydning, J., The digitization of the world from edge to core. \textit{IDC} \textbf{16}, 1--28 (2018).

\bibitem{gu2014optical}Gu, M., Li, X., Cao, Y., Optical storage arrays: a perspective for future big data storage. \textit{Light Sci. Appl.} \textbf{3}, e177 (2014).

\bibitem{levandoski2013identifying}Levandoski, J., Larson, P., Stoica, R., Identifying hot and cold data in main-memory databases. \textit{ICDE.}, 26-37 (2013).

\bibitem{fleischer2020cooling}Fleischer, A., Cooling our insatiable demand for data. \textit{Science} \textbf{370}, 783--784 (2020).

\bibitem{zhang2014seemingly}Zhang, J., Gecevi{\v{c}}ius, M., Beresna, M., Kazansky, P., Seemingly unlimited lifetime data storage in nanostructured glass. \textit{Phys. Rev. Lett.} \textbf{112}, 033901 (2014).

\bibitem{gu2010road}Gu, M., Li, X., The road to multi-dimensional bit-by-bit optical data storage. \textit{Optics and Photonics News} \textbf{21}, 28--33 (2010).


\bibitem{stiller2017cascaded}Stiller, B., \textit{et al.} Cascaded waveguide-based photon-phonon memory. \textit{CLEO/Europe-EQEC}, 1-1 (2017).

\bibitem{seddighian2008optical}Seddighian, P., Optical packet switching using multi-wavelength labels. \textit{Citeseer} (2008).

\bibitem{meiling2019research}Ling, M., Zhang, M., Li, X., Cao, Y., Research progress of super-resolution optical data storage. \textit{Opto-Electronic Engineering} \textbf{46}, 180649--1 (2019).

\bibitem{wei2013dynamic}Wei, J., On the dynamic readout characteristic of nonlinear super-resolution optical storage. \textit{Opt. Commun.} \textbf{291}, 143--149 (2013).

\bibitem{mansuripur1997principles}Mansuripur, M., Sincerbox, G., Principles and techniques of optical data storage. \textit{Proc. IEEE} \textbf{85}, 1780--1796 (1997).

\bibitem{okamoto2011holographic}Okamoto, A., Kunori, K., Takabayashi, M., Tomita, A., Sato, K., Holographic diversity interferometry for optical storage. \textit{Opt. Express} \textbf{19}, 13436--13444 (2011).

\bibitem{hvilsted2009volume}Hvilsted, S., S{\'a}nchez, R. Alcal{\'a}, C., The volume holographic optical storage potential in azobenzene containing polymers. \textit{J. Mater. Chem.} \textbf{19}, 6641--6648 (2009).

\bibitem{glezer1996three}Glezer, E. \textit{et al.} Three-dimensional optical storage inside transparent materials. \textit{Opt. Lett.} \textbf{21}, 2023--2025 (1996).

\bibitem{qiu1998three}Qiu, J., Miura, K., Inouye, H., Nishii, J., Hirao, K., Three-dimensional optical storage inside a silica glass by using a focused femtosecond pulsed laser. \textit{Nucl Instrum Methods Phys Res B} \textbf{141}, 699--703 (1998).

\bibitem{zhang2016eternal}Zhang, J., Eternal 5D data storage by ultrafast laser writing in glass. \textit{Laser-based Micro-and Nanoprocessing X} \textbf{9736}, 163--178 (2016).

\bibitem{bhardwaj2006optically}Bhardwaj, V. \textit{et al.} Optically produced arrays of planar nanostructures inside fused silica. \textit{Phys. Rev. Lett.} \textbf{96}, 057404 (2006).

\bibitem{lei2023ultrafast}Lei, Y. \textit{et al.} Ultrafast Laser Writing in Different Types of Silica Glass. \textit{Laser Photonics Rev.}, 2200978 (2023).

\bibitem{shimotsuma2003self}Shimotsuma, Y., Kazansky, P., Qiu, J., Hirao, K., Self-organized nanogratings in glass irradiated by ultrashort light pulses. \textit{Phys. Rev. Lett.} \textbf{91}, 247405 (2003).



\bibitem{shimotsuma2010ultrafast}Shimotsuma, Y. \textit{et al.} Ultrafast Manipulation of Self-Assembled Form Birefringence in Glass. \textit{Adv Mater} \textbf{22}, 4039--4043 (2010).

\bibitem{sakakura2020ultralow}Sakakura, M., Lei, Y., Wang, L., Yu, Y., Kazansky, P., Ultralow-loss geometric phase and polarization shaping by ultrafast laser writing in silica glass. \textit{Light Sci. Appl.} \textbf{9}, 15 (2020).

\bibitem{wang2022100}Wang, H. \textit{et al.} 100-Layer error-free 5D optical data storage by ultrafast laser nanostructuring in glass. \textit{Laser Photonics Rev.} \textbf{16}, 2100563 (2022).



\bibitem{lei2021high}Lei, Y. \textit{et al.} High speed ultrafast laser anisotropic nanostructuring by energy deposition control via near-field enhancement. \textit{Optica} \textbf{8} 1365--1371 (2021).

\bibitem{yang2023high}Yang, D. \textit{et al.} High optical storage density using three-dimensional hybrid nanostructures based on machine learning. \textit{Opt Lasers Eng} \textbf{161}, 107347 (2023).

\bibitem{milster1999fundamental}Milster, T., Upton, R., Fundamental principles of crosstalk in optical data storage. \textit{Jpn J Appl Phys} \textbf{38}, 1608 (1999).

\bibitem{yang1999interlayer}Yang, B., Shieh, H., Interlayer cross talk in dual-layer read-only optical disks. \textit{Appl. Opt.} \textbf{38}, 333--338 (1999).

\bibitem{yi1995statistical}Yi, X., Campbell, S., Yeh, P., Gu, C., Statistical analysis of cross-talk noise and storage capacity in volume holographic memory: image plane holograms. \textit{Opt. Lett.} \textbf{20}, 779--781 (1995).

\bibitem{gao20214d}Gao, L., Zhang, Q., Evans, R., Gu, M., 4D Ultra-High-Density Long Data Storage Supported by a Solid-State Optically Active Polymeric Material with High Thermal Stability. \textit{Adv. Opt. Mater.} \textbf{9}, 2100487 (2021).

\bibitem{lamon2021nanophotonics}Simone, L., Zhang, Q., Gu, M., Nanophotonics-enabled optical data storage in the age of machine learning. \textit{APL Photonics} \textbf{6}, 11 (2021).

\bibitem{van2006cancellation}Beneden, S., Riani, J., Bergmans, J., Immink, A., Cancellation of linear intersymbol interference for two-dimensional storage systems. \textit{IEEE Trans. Magn.} \textbf{42}, 2096--2106 (2006).




\bibitem{dong2011estimating}Dong, G., Pan, Y., Xie, N., Varanasi, C., Zhang, T., Estimating information-theoretical NAND flash memory storage capacity and its implication to memory system design space exploration. \textit{IEEE Trans Very Large Scale Integr VLSI Sys} \textbf{20}, 1705--1714 (2011).


\bibitem{wang2011error}Wang, X., Dong, G., Pan, L., Zhou, R., Error correction codes and signal processing in flash memory. \textit{Flash Memories} 57--82 (2011).

\bibitem{zhang2022low}Zhang, X., Yung, M., Low-depth optical neural networks. \textit{Chip} \textbf{1}, 100002 (2022)

\bibitem{wicker1999reed}Wicker, S., Bhargava, V., \textit{Reed-Solomon codes and their applications} (IEEE Press, 1999).

\bibitem{zheng2007surface}Zheng, L., Schmid, A., Lambropoulos, J., Surface effects on Young’s modulus and hardness of fused silica by nanoindentation study. \textit{J. Mater. Sci.} \textbf{42}, 191--198 (2007).

\bibitem{penn2006frequency}Penn, S. \textit{et al.} Frequency and surface dependence of the mechanical loss in fused silica. \textit{Phys. Lett. A} \textbf{352}, 3--6 (2006).

\bibitem{xiao2021models}Xiao, H. \textit{et al.} Models of grinding-induced surface and subsurface damages in fused silica considering strain rate and micro shape/geometry of abrasive. \textit{Ceram. Int.} \textbf{47}, 24924--24941 (2021).

\bibitem{sudrie1999writing}Sudrie, L., Franco, M., Prade, B., Mysyrowicz, A., Writing of permanent birefringent microlayers in bulk fused silica with femtosecond laser pulses. \textit{Opt. Commun.} \textbf{171}, 279--284 (1999).

\bibitem{mills2002embedded}Mills, J., Kazansky, P., Bricchi, E., Baumberg, J., Embedded anisotropic microreflectors by femtosecond-laser nanomachining. \textit{Appl. Phys. Lett.} \textbf{81}, 196--198 (2002).

\bibitem{latychevskaia2019lateral}Latychevskaia, T., Lateral and axial resolution criteria in incoherent and coherent optics and holography, near-and far-field regimes. \textit{Appl. Opt.} \textbf{58}, 3597--3603 (2019).

\bibitem{zheludev2008diffraction}Zheludev, N., What diffraction limit? \textit{Nat. Mater} \textbf{7}, 420--422 (2008).

\bibitem{fischer2013three}Fischer, J., Wegener, M., Three-dimensional optical laser lithography beyond the diffraction limit. \textit{Laser Photonics Rev.} \textbf{7}, 22--44 (2013).

\bibitem{taylor2008applications}Taylor, R., Hnatovsky, C., Simova, E., Applications of femtosecond laser induced self-organized planar nanocracks inside fused silica glass. \textit{Laser Photonics Rev.} \textbf{2}, 26--46 (2008).


\bibitem{kogelnik1965propagation}Kogelnik, H., On the propagation of Gaussian beams of light through lenslike media including those with a loss or gain variation. \textit{Appl. Opt.} \textbf{4}, 1562--1569 (1965).

\bibitem{mccall1969self}McCall, S., Hahn, E., Self-induced transparency. \textit{Phys. Rev.} \textbf{183}, 457 (1969).

\bibitem{belkina2019automated}Belkina, A. \textit{et al.} Automated optimized parameters for T-distributed stochastic neighbor embedding improve visualization and analysis of large datasets. \textit{Nat. Commun} \textbf{10} 5415 (2019).

\bibitem{kanal1978models}Kanal, L., Sastry, A., Models for channels with memory and their applications to error control. \textit{Proc. IEEE} \textbf{66}, 724--744 (1978).

\bibitem{shrivastava2013error}Shrivastava, P., Singh, U., Error detection and correction using Reed Solomon codes. \textit{Int. j. adv. res. comput. sci. softw. eng.} \textbf{3} 8 (2013).

\bibitem{yang2023deep}Yang, J., Cui, K. \textit{et al.} Deep-learning based on-chip rapid spectral imaging with high spatial resolution. \textit{Chip} \textbf{2}, 100045 (2023).


\bibitem{lecun2015deep}Cun, Y., Bengio, Y., Hinton, G., Deep learning. \textit{Nature} \textbf{521}, 436--444 (2015).

\bibitem{tan2021photonic}Tan, D., Wang, Z., Xu, B., Qiu, J., Photonic circuits written by femtosecond laser in glass: improved fabrication and recent progress in photonic devices. \textit{Adv. Photonics} \textbf{3}, 024002--024002 (2021).



\bibitem{gu2016nanomaterials} Gu, M., Zhang, Q., Simone, L., Nanomaterials for optical data storage. \textit{Nature Reviews Materials} \textbf{1}, 1--14 (2016).


\end{thebibliography}

\end{document}